# Ni modified Fe$_3$O$_4$(001) surface as a simple model system for understanding the Oxygen Evolution Reaction

Francesca Mirabella[+,1,#], Matthias Müllner[+,1], Thomas Touzalin[+,2], Michele Riva[1], Zdenek Jakub[1], Florian Kraushofer[1], Michael Schmid[1], Marc T.M. Koper[2], Gareth S. Parkinson[1], Ulrike Diebold[1,*]

[1]*Institut für Angewandte Physik, Technische Universität Wien, A-1040 Wien, Austria*
[2]*Leiden Institute of Chemistry, Leiden University, PO Box 9502, 2300 RA, Leiden, The Netherlands*

[+] = These authors contributed equally.
[*] = corresponding author.
[#] = current address: Bundesanstalt für Materialforschung und -prüfung, Unter den Eichen 44-46, 12203 Berlin, Germany

**Abstract**

Electrochemical water splitting is an environmentally friendly technology to store renewable energy in the form of chemical fuels. Among the earth-abundant first-row transition metal-based catalysts, mixed Ni-Fe oxides have shown promising performance for effective and low-cost catalysis of the oxygen evolution reaction (OER) in alkaline media, but the synergistic roles of Fe and Ni cations in the OER mechanism remain unclear. In this work, we report how addition of Ni changes the reactivity of a model iron oxide catalyst, based on Ni deposited on and incorporated in a magnetite Fe$_3$O$_4$ (001) single crystal, using a combination of surface science techniques in ultra-high-vacuum such as low energy electron diffraction (LEED), x-rays photoelectron spectroscopy (XPS), low energy ion scattering (LEIS), and scanning tunneling microscopy (STM), as well as atomic force microscopy (AFM) in air, and electrochemical methods such cyclic voltammetry (CV) and electrochemical impedance spectroscopy (EIS) in alkaline media. A significant improvement in the OER activity is observed when the top surface presents an Fe:Ni composition ratio in the range 20-40%, which is in good agreement with what has been observed for powder catalysts. Furthermore, a decrease in the OER overpotential is observed following surface aging in electrolyte for three days. At higher Ni load, AFM shows the growth of a new phase attributed to an (oxy)-hydroxide phase which, according to CV measurements, does not seem to correlate with the surface activity towards OER. EIS suggests that the OER precursor species observed on the clean and Ni-modified surfaces are similar and Fe-centered, but form at lower overpotentials when the surface Fe:Ni ratio is optimized. We propose that the well-defined Fe$_3$O$_4$(001) surface can serve as a model system for understanding the OER mechanism and establishing the structure-reactivity relation on mixed Fe-Ni oxides.

**Introduction**

Currently most energy sources used by our society are based on fossil fuels. Their combustion (coal, oil, and gas), together with large-scale deforestation, is causing massive emissions of greenhouse gases. Given the destructive environmental impact of these gases, effort has focused on the production, storage and transport of renewable energy (wind or sunlight)[1]. A promising technology to address this issue uses renewable energy to produce chemical energy through the splitting of water into hydrogen and oxygen (water electrolysis)[2]. However, the efficiency of the electrolysis process is hampered by the sluggish kinetics of water oxidation to $O_2$, also known as oxygen evolution reaction (OER). This reaction has been described as the bottleneck of the water splitting and understanding its mechanism at the atomic scale could be a first step in addressing this challenge[2]. Many catalysts have been proposed to reduce the overpotential losses for OER and investigated in different pH conditions[3], from acidic ($2H_2O \rightarrow 4H^+ + O_2 + 4e^-$) to alkaline ($4OH^- \rightarrow 2H_2O + O_2 + 4e^-$) media. In acidic media, noble metals such as Ru or Ir show promising OER stability and activity. However, due to their limited availability and high price many researchers are seeking alternative catalysts based on earth-abundant elements[3,4,5,6]. In an alkaline environment, oxides and hydroxides of late first-row transition metals (Mn, Fe, Co, Ni) have been found to have comparable performances to noble metals[3]. In particular, NiFe-based (oxy)hydroxide catalysts are reported to show the lowest overpotential for OER in alkaline conditions (pH 13 and 14)[7], but the synergistic role of Fe and Ni is still under debate.

Comparing OER catalysts is complicated by many underlying factors, including differences in electrochemically active surface area, catalyst electrical conductivity, surface chemical stability, surface composition, and reaction mechanism. In this work, we describe our efforts to circumvent these issues by using a combined surface science/electrochemistry approach. We have prepared well-defined Ni-modified $Fe_3O_4(001)$ surfaces in ultra-high-vacuum (UHV) with different Fe:Ni ratios and, after characterization with surface science techniques, we have studied their electrochemical performances towards OER using cyclic voltammetry and electrochemical impedance spectroscopy. A significant increase in the OER activity is observed as the Ni content increases, and the optimum composition has ratios of Fe:Ni in the top surface layer in the range of 20-40%. These results are in good agreement with literature for the best OER powder catalysts[7]. Furthermore, based on the analysis of the surface morphological changes before and after reaction, together with adsorption capacitance measurements, we propose that the active sites responsible for the formation of the OER precursor are the same on the clean and on the Ni-modified magnetite. Nevertheless, the presence of the Ni on the surface shifts the formation of this precursor to lower overpotential.

Our study provides a well-defined model catalyst that is at the same time simple, highly active, and stable under operation conditions, and therefore ideal to be used as model system to gain atomic scale insights into the complicated OER mechanism.

1. **Experimental Details**

**UHV preparation and characterization.** The experiments were performed on a natural $Fe_3O_4$(001) single crystal (SurfaceNet GmbH) prepared in UHV by cycles of 1 keV $Ar^+$ sputtering and 900 K annealing. Every other annealing cycle was performed in an $O_2$ environment ($p_{O_2}$ = 5×10$^{-7}$ mbar, 20 min) to maintain the stoichiometry of the crystal selvedge. Surface analysis was performed in a UHV system with a base pressure <10$^{-10}$ mbar, furnished with a commercial Omicron SPECTALEED rear-view optics and an Omicron UHV STM-1. XPS data were acquired using non-monochromated Al Kα x-rays and a SPECS PHOIBOS 100 electron analyser at grazing emission (70° from the surface normal). The same analyser was used to carry out the low-energy $He^+$ ion scattering (LEIS) experiments (1.225 keV $He^+$, scattering angle 137°), an exquisitely surface-sensitive technique. Ni was deposited using a Focus electron-beam evaporator, for which the deposition rates were calibrated using a temperature-stabilized quartz crystal microbalances (QCM). One monolayer (ML) is defined as one atom per (√2 × √2)R45° unit cell, which corresponds to 1.42 × 10$^{14}$ atoms/cm$^2$. Ni depositions higher than 2 ML were prepared by first depositing 2 ML Ni on the surface at room temperature, followed by mild annealing at 200°C for 10 min. This causes a transition from Ni being present as 2-fold coordinated adatoms to 6-fold coordinated "incorporated" cations[8], see Fig. 1; the procedure was then repeated as many times as necessary to reach the desired coverage.

**Characterization in-air.** After UHV-preparation and characterization as well as after the electrochemical measurements, the samples were brought to air and imaged using an Agilent 5500 ambient AFM in intermittent contact mode with Si tips on Si cantilevers.

**Electrochemical measurements**. Cyclic voltammetry and impedance spectroscopy were performed using a Metrohm-Autolab PGSTAT32 potentiostat and a custom-made electrochemical flow cell (made from perfluoroalkoxy alkane, PFA), mounted to the vacuum chamber. Prior to experiments, the chamber was filled with Ar (99.999%, Air Liquide, additionally purified with Micro-Torr point-of-use purifiers, SAES MC50–902 FV) to ambient pressure. The contact between sample and flow cell was sealed with Kalrez O-rings. Prior to measurements, the electrolyte reservoir was evacuated and ultrasonicated to remove dissolved $CO_2$. The flow cell was filled with electrolyte by increasing the pressure in the electrolyte compartment with Ar to slight overpressure. A glassy carbon counter electrode and a leak-free Ag/AgCl reference electrode (Innovative Instruments Inc.) were used. For impedance measurements, the latter was coupled to a glassy carbon quasi-reference electrode through a 100 nF capacitor. All electrochemical data were corrected for $iR_u$ drop; the uncompensated solution resistance $R_u$ was determined from impedance Nyquist plots by extrapolating the minimum total impedance in the linear regime between 10 kHz and 100 kHz. All electrochemical potentials are referred to either the measured Ag/AgCl reference electrode $E_{Ag/AgCl}$ or given as the overpotential η, which was determined via the equation η = $E_{Ag/AgCl}+E_{RHE}-1.229-iR_u$. $E_{RHE}$ is the potential of the reversible hydrogen electrode (RHE) vs a Ag/AgCl electrode. The potential of the RHE (Hydroflex) was measured before and after the electrochemical measurements to improve consistency of the results. The electrolyte was prepared from level-1 water (Merck Milli-Q, ρ= 18.2 MΩ·cm, 3 ppb total organic carbon), and reagent-grade NaOH (50 mass % in water, Sigma-Aldrich). Prior to use, all glassware and PFA parts where cleaned by boiling in 20% nitric acid and copious rinsing with Milli-Q water.

2. Results

3.1 Characterization of the Catalyst Surface before Reaction

Figure 1a shows a schematic model of the UHV-prepared $Fe_3O_4$(001) surface. The surface is oxidized with respect to the bulk $Fe_3O_4$ and is not a simple bulk truncation. Specifically, an interstitial tetrahedrally coordinated iron in the second layer ($Fe_{tet}$, light blue in the model) replaces two octahedrally coordinated iron atoms ($Fe_{oct}$, dark blue) in the third layer[9], giving rise to a (√2 × √2)R45° periodicity. All surface Fe is in the 3+ state in the so-called subsurface cation vacancy (SCV) reconstruction, and it is the most stable termination of $Fe_3O_4$(001) over the range of oxygen chemical potentials encountered in UHV-based experiments[9].

In the lower part of Figure 1a, a typical STM image of the UHV-prepared $Fe_3O_4$(001) surface is shown. Undulating rows of surface Fe atoms appearing as protrusions run in the [110] direction. It is common to observe surface hydroxyl groups $O_sH$ (i.e. hydrogen atoms bonding to surface oxygen atoms, which are themselves not imaged) as bright protrusions on the Fe rows. This occurs because the hydroxyl modifies the density of states of the nearby Fe cations, causing them to appear brighter in empty-states STM images[10,11]. Figure 1a also displays other common defects visible on the clean surface, such as antiphase domain boundaries, which are imaged as meandering line defects, and unreconstructed unit cells, which appear similar to two neighboring hydroxyl groups. These are caused by two additional Fe atoms in the subsurface layer (instead of one interstitial Fe), which again modifies the density of states of the surface atoms[10,12]. It is not possible to image the surface oxygen in STM as they have no density of states in the vicinity of the Fermi level. However, their positions are exactly known from density functional theory calculations and quantitative low-energy electron diffraction (red in model in Figure 1a)[9].

The surface reconstruction makes it possible to progressively modify the magnetite surface and accommodate foreign metal atoms (such as nickel) in specific positions.[8] Following Ni evaporation under the appropriate temperature conditions, it is possible to obtain two different Ni geometries: Ni adatoms 2-fold coordinated to surface oxygen atoms (model in Figure 1b, green) and incorporated Ni occupying octahedrally coordinated sites below the surface (model in Figure 1c)[8,13]. Ni deposition at room temperature leads to Ni adatoms in the 2-fold coordination, which are imaged in STM as isolated, bright protrusions appearing between the Fe rows (light blue circles in Fig. 1b). The transition from 2-fold to 6-fold coordination is achieved by annealing the surface at 200 °C for 10 minutes. As the incorporated Ni are in the subsurface, they cannot be imaged directly in STM, but they modify the electronic structure of the nearby Fe cations, making them to appear brighter in empty-state images (red circles in Figure 1c)[8,13].

Their appearance is similar to the unreconstructed cell discussed earlier (Figure 1a). Furthermore, the STM image in Figure 1c shows additional protrusions within the Fe rows (highlighted with yellow circles), which we previously assigned to Ni replacing Fe atoms in the 5-fold-coordinated position in the top-surface layer[13].

The incorporation of Ni in the vacant subsurface octahedral site is only possible if the interstitial $Fe_{tet}$ moves back into the other subsurface octahedral site of the unit cell. The resulting cation rearrangement closely resembles a bulk-truncated $Fe_3O_4$(001) surface[8,14], and a (1 × 1) periodicity is observed in LEED. It is possible to recover clean (√2 × √2)R45° reconstructed surface by annealing to high temperatures, which causes the Ni atoms to diffuse into deep bulk layers.

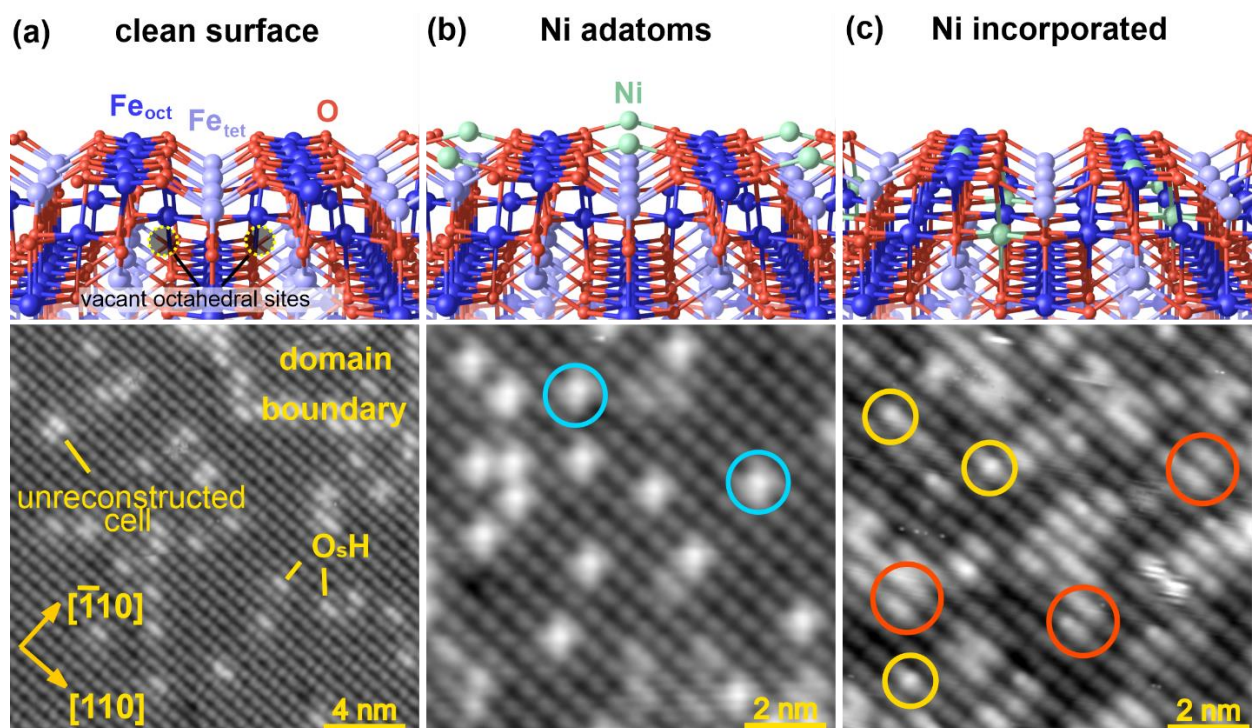

*Figure 1. Atomic models showing side views of the (a) $Fe_3O_4$(001) clean, (b) doped with 2-fold coordinated Ni adatoms, and (c) with 6-fold coordinated Ni incorporated, as well as corresponding STM images.* The clean $Fe_3O_4$(001) surface in (a) exhibits a (√2 × √2)R45° reconstruction due to subsurface cation vacancies (SCV) (octahedrally coordinated $Fe_{oct}$ are dark blue, tetrahedrally coordinated $Fe_{tet}$ are light blue, O atoms are red, and Ni atoms are light green). The STM image below shows the clean surface and its common defects: surface $O_sH$ groups, antiphase domain boundaries and unreconstructed unit cells. In (b) Ni adatoms are adsorbed in the 2-fold-coordinated surface sites, each formed by two undercoordinated O atoms. In the STM image below the Ni adatoms are highlighted by light blue circles (the Ni coverage is 0.15 ML). In (c) the Ni is incorporated into the subsurface and occupies the vacant octahedral site or replaces a surface $Fe_{oct}$ atom in a metastable configuration. Both types of incorporated Ni are shown in the STM image below, highlighted by red and yellow respectively. (The coverage of the Ni adatoms was 0.40 Monolayers (ML) before the incorporation, where 1 ML Ni is defined as one nickel atom per (√2 × √2)R45° unit cell or $1.42 \times 10^{14}$ atoms per $cm^2$. This figure is adapted from Ref.13.

Hereafter, we deal exclusively with the incorporated Ni-doped magnetite shown in Fig. 1c, which resembles the structure of mixed spinel ferrite, i.e., a $Ni_xFe_{3-x}O_4$-like system, suggested to be one of the active phases in OER[15,16].

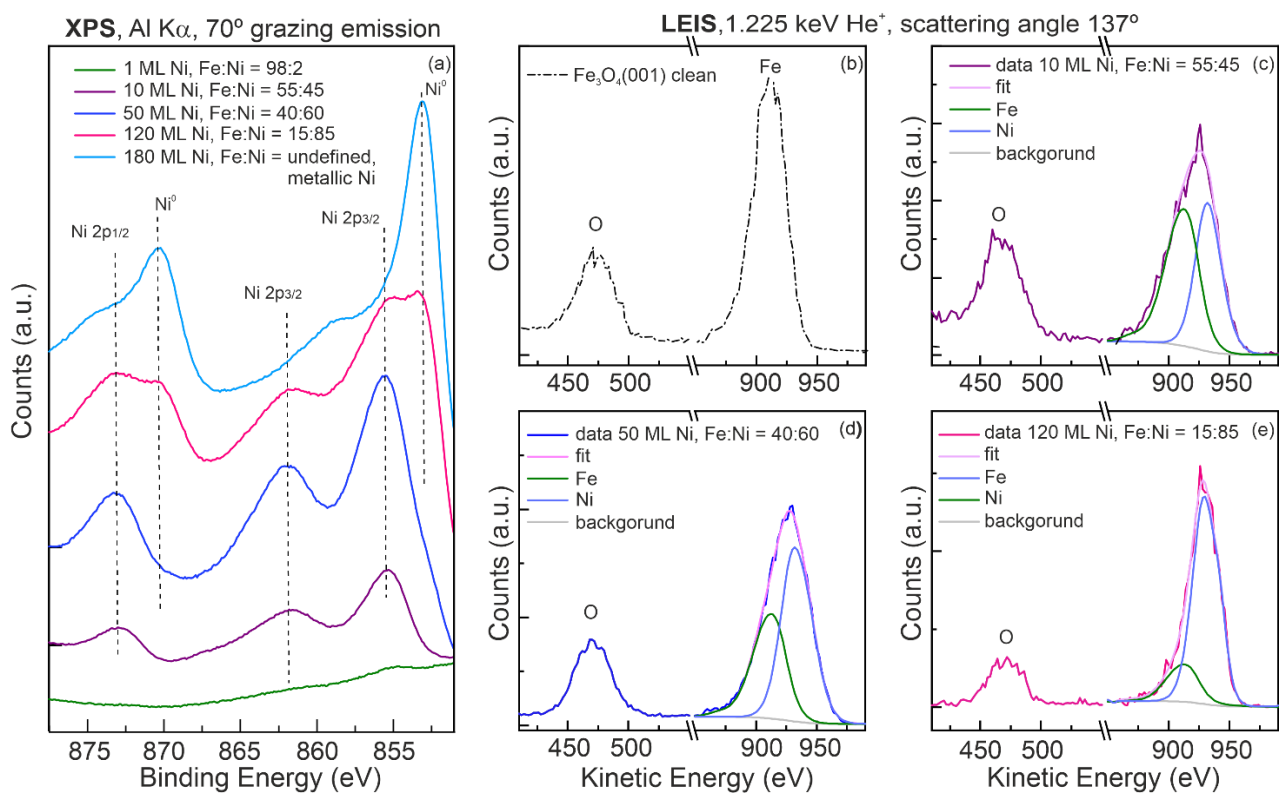

*Figure 2. XPS (Al Kα, 70° grazing emission) and LEIS spectra (1.225 keV He$^+$, scattering angle 137°).* (a) XPS spectra of the Ni 2p region after doping the $Fe_3O_4(001)$-(√2 × √2)R45° surface with 1 ML Ni (green), 10 ML Ni (purple), 50 ML Ni (blue), 120 ML Ni (pink), and 180 ML Ni (light blue). (b)-(e) LEIS spectra of the clean and Ni-doped $Fe_3O_4(001)$-(√2 × √2)R45° surfaces shown in (a), and corresponding fits. The fitted spectra show an Fe:Ni ratio of (c) 55:45, (d) 40:60, and (e) 15:85, corresponding to 10 ML Ni-, 50 ML Ni-, and 120 ML Ni-doped magnetite surface respectively. The LEIS spectra shown are the result of the average of several consecutive scans and their fits were obtained using the software CasaXPS.

The XPS spectra in Figure 2a shows the Ni *2p* region for different coverages after Ni was deposited onto the $Fe_3O_4(001)$ surface at room temperature and annealed at 200°C. Five different total Ni depositions are considered: 1 ML (green), 10 ML (purple), 50 ML (blue), 120 ML (pink), and 180 ML (light blue).

After deposition of 1 ML, a small signal is observed in XPS at 855.5 eV, corresponding to the Ni *2p$_{3/2}$* peak[17,7]. This is a higher binding energy than metallic Ni[17], which, together with the strong satellite at ≈862 eV, indicates that the nickel is oxidized. Earlier DFT calculations predicted that incorporated Ni atoms are Ni(II)[8], as in $NiFe_2O_4$.

As the Ni deposition increases to 10 ML, the Ni $2p_{3/2}$ at 855.5 eV increases in intensity, together with the 861.9 eV and the $2p_{1/2}$ peak at 873 eV, which are harder to see at lower Ni coverage. These features increase in intensity as the Ni deposition increases up to 50 ML. At even higher Ni load (120 ML), two new signals at 853.1 eV and 870.2 eV emerge, indicating that metallic Ni is present on the surface[17]. Above 120 ML, the XPS spectrum changes shape to a peak with only two main features at 853.1 eV and 870.2 eV, indicating that the surface is fully covered with metallic Ni.

We imaged the $Fe_3O_4$(001) surface before and after Ni-doping using ambient AFM right after removing the crystal from the UHV chamber (Figure 3a-d). The corresponding LEED patterns acquired in UHV are shown as insets in each AFM image.

The clean $Fe_3O_4$(001) surface appears overall flat in ambient AFM, with micrometer-wide terraces separated by step bunches[18] (Figure 3a). The corresponding LEED pattern exhibits the (√2 × √2)R45° periodicity of the SCV reconstruction[9] (yellow square in the inset).

Figure 3b shows the AFM image of a magnetite surface doped with 50 ML Ni. The large terraces as well as the step bunches observed earlier[18] on the clean magnetite remain visible, suggesting that the doping did not affect the overall surface morphology. Isolated (white) features 0.4-0.6 nm high are visible on the surface. Based on the corresponding line profile (Figure 3a´´, blue), which shows step heights similar to what is observed in Figure 3a, we suspect these to be residues originating from dust or carbonaceous species. The LEED pattern in the inset shows that the reconstruction spots are now absent and a (1 × 1) symmetry is observed (blue square), which is known to occur above 1 ML Ni atoms incorporated in the subsurface[8].

Figure 3c-d show AFM images of magnetite surfaces following doping with 120 ML and 180 ML Ni, respectively. The surface in (c) exhibits a rougher morphology than observed in (a) and (b), with a corrugation of ≈0.5 nm (Figure 3c´´, lilac). Accordingly, the corresponding LEED pattern shows weaker (1 × 1) spots. Following higher Ni doping, the surface morphology changes considerably (d). Although the step bunches are still visible underneath, the surface appears covered in round features having height of ~2nm (Figure 3d´´, lilac). Based on the XPS data showed in Figure 2a, we assign these features to metallic Ni clusters. The corresponding LEED pattern shows very weak (1 × 1) spots with a high background, indicating an increasing fraction of the surface covered by structures with no well-defined crystallographic relationship to the substrate, in agreement with the presence of metallic agglomerates on the surface.

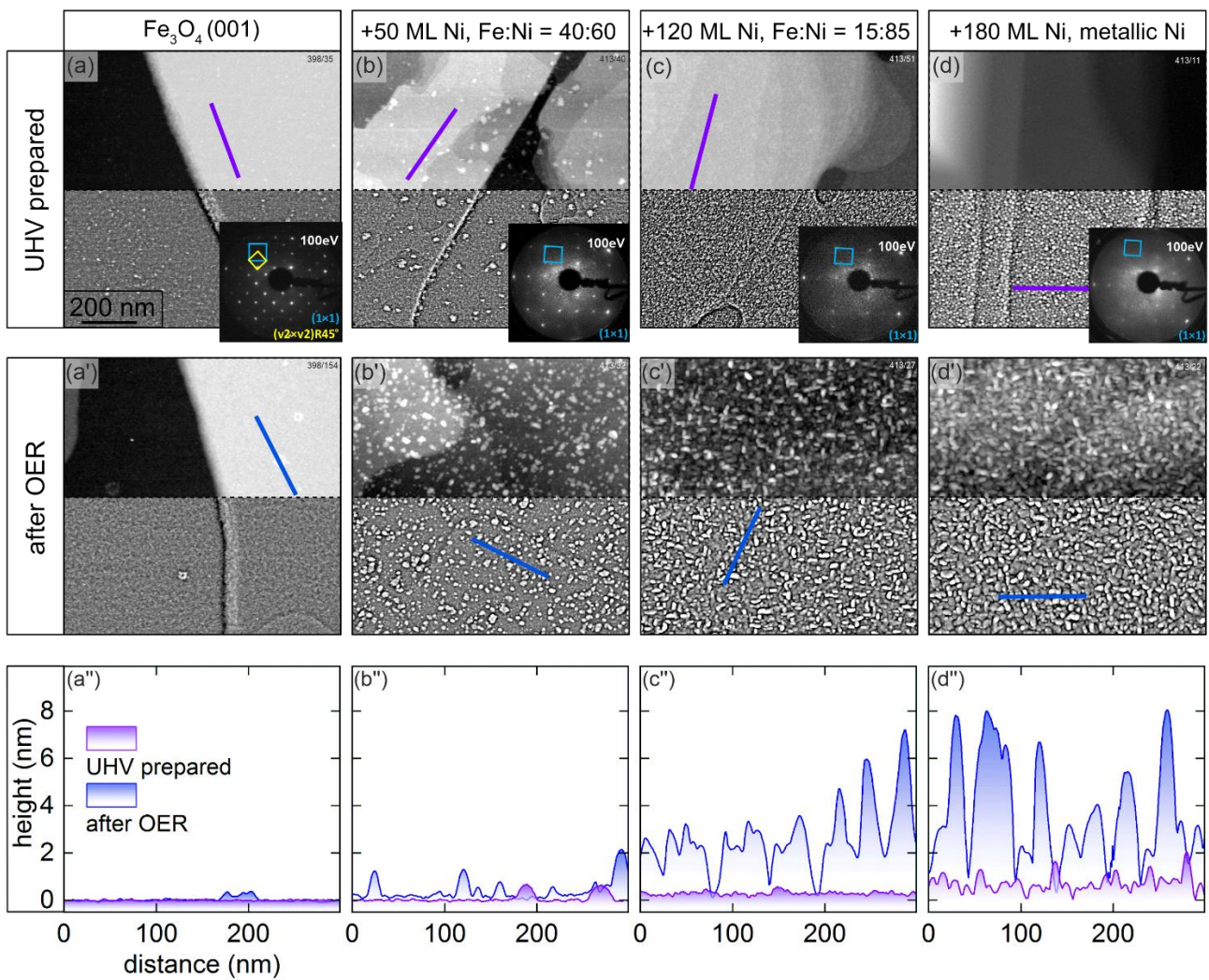

*Figure 3. Ambient-AFM images of the clean and Ni-doped Fe₃O₄ (001) surfaces before and after OER in 1 M NaOH.* The surface morphologies of the (a) clean Fe₃O₄(001), and Ni-doped surfaces following (b) 50 ML Ni, (c) 120 ML Ni, and (d) 180 ML Ni deposition before OER are shown together with the corresponding LEED pattern acquired in UHV (insets). The middle part of the figure (a´-d´) shows the morphology of the above-mentioned surfaces after they have been exposed to the electrolyte for three days and cycled until a stable OER current was observed. Each AFM image is shown without (top half) and with (bottom half) high-pass filter. The bottom part of the figure, (a´´-d´´), shows the corresponding line profiles of the surfaces as prepared in UHV (lilac) and after OER (blue).

A quantitative measurement of the surface composition, given as the Fe:Ni ratio for each Ni modified surface can be obtained with LEIS measurements (Figure 2b-e). The clean surface exhibits a LEIS peak centered at 910 eV (Figure 2b), corresponding to the surface Fe atoms. Following 10 ML Ni-doping, the LEIS signal is broader and shifts to higher kinetic energy KE (Figure 2c, purple). This peak can be well fitted by a (slightly shifted) peak from the surface Fe and an additional component at 931 eV corresponding to the Ni (Figure 2c, green and purple respectively). The fit allows us to estimate an Fe:Ni top surface ratio on the 10 ML Ni-doped surface of 55:45. Similarly, we calculate that the surfaces following 50 ML and 120 ML Ni-doping show Fe:Ni ratios of 40:60 and 15:85, respectively. At higher Ni-doping (180 ML) the whole surface results covered in metallic Ni particles, which makes it difficult to use LEIS to quantify the Fe:Ni surface ratio. Therefore, we restrict ourselves to the coverage regime prior to the formation of metallic Ni clusters.

Figure 2b-e also shows how the surface oxygen peak (centered at ~470 eV) evolves as a function of the Ni doping. The intensity of the surface oxygen peak seems to remain constant as the Fe:Ni ratio decreases down to 40:60. Differently, a clear decrease in the oxygen intensity is observed for the surface with lower Fe:Ni ratio (15:85). We can speculate that this behavior correlates with the presence of some metallic Ni on top, as observed in the XPS in Figure 2a, pink.

Importantly, no systematic change in consecutive scans was observed, which rules out substantial damage to the surface by He$^+$ sputtering during LEIS measurements. In what follows, we will use the LEIS-determined Fe:Ni ratio to refer to our model catalysts.

## 2.2 Electrochemical Performance

The electrochemical performance of the clean and Ni-doped $Fe_3O_4$(001) surfaces was investigated using cyclic voltammetry. The overpotential required to reach a given current density is a key catalytic parameter to compare several catalysts and to estimate the energetic efficiency of integrated (photo-) electrochemical water splitting devices[3]. The cyclic voltammograms (Figure 4a) were acquired in 1 M NaOH under Ar with a scan rate of 10 mV s$^{-1}$ after cycling the electrode until a stable OER current could be observed on two subsequent CVs. Data corresponding to the surfaces imaged in Figure 3a-d are shown, as well as for surfaces with a Fe:Ni ratio of 98:2 and 55:45.

The clean $Fe_3O_4$(001) surface shows an overpotential of 597 mV at a current density of 5 mA·cm$^{-2}$ (Figure 4a, black), and the surface with a Fe:Ni ratio of 98:2 (green) exhibits similar performance. As the Ni content in the subsurface increases, higher activity towards OER is observed. The OER overpotential decreases by ~110 mV when the Fe:Ni ratio is 55:45 (purple), and reaches ~340 mV vs RHE when the Fe:Ni ratio is 40:60. A higher Ni load (Fe:Ni = 15:85, pink) results in a similar activity as the surface with Fe:Ni ratio of 40:60. Additionally, the surface with a Fe:Ni ratio of 15:85 exhibits a pair of anodic and cathodic peaks at 1.369 and 1.311 V *vs* RHE respectively (pink, inset in figure 4a), consistent with the reversible oxidation of Ni(II) to a higher oxidation state (III), as it is reported for the case of the nickel hydroxide/oxyhydroxide couple (Ni(OH)$_2$/NiOOH)[19]. It can also be observed that the charge (peak area) corresponding to this peak increases with cycling, indicating the growth of a thicker Ni oxide film on top of the $Fe_3O_4$(001) surface. These observations suggest a change in the Fe:Ni ratio at the surface following electrode cycling. When only metallic Ni is present on the as-prepared sample, an increase in the overpotential of ~88 mV is observed (Figure 4a, light blue). A corresponding increase in the charge of the Ni(OH)$_2$/NiOOH peak is observed, as well as anodic shifts of 170 mV and 130 mV for the anodic and cathodic peaks respectively. A similar anodic shift of the Ni peak has been observed with increasing Fe:Ni ratio in the NiOOH phase either by co-deposition of Fe during the film synthesis[20,21] or by incorporation of Fe impurities from the electrolyte into NiOOH electrodes[22]. The NiOOH phase formed from metallic Ni at the surface of the as-prepared surface incorporates more Fe than in the surface with a Fe:Ni ratio of 15:85. Moreover, the charge of the Ni(OH)$_2$/NiOOH peak remains constant with cycling, indicating a saturation of the surface with nickel (oxy)hydroxide.

As a comparative metric of activity, Tafel plots are also shown (Figure 4b). The determination of Tafel slopes can help elucidating the rate limiting step of a mechanism, but their analysis is particularly difficult in the case of multiple electron-proton transfer reactions such as OER[3]. The clean and low Ni-doped (Fe:Ni

= 98:2) surfaces display values of 92 mV/dec and 88 mV/dec respectively, whereas the Ni-doped $Fe_3O_4$(001) surfaces with a Ni load of 50-85% all show similar values in the range 50-61 mV/dec.

In Figure 4c we plot the overpotential values and the Tafel slopes showed in Figure 4a-b as a function of the surface Fe:Ni ratios. Interestingly, the lowest OER overpotential values are obtained for the catalysts with a surface Fe:Ni ratio between 15-40 %, in agreement with what is reported in literature for the best OER powder catalysts[7]. Furthermore, the Tafel slopes values fall in the same range as observed for NiFe (oxy)hydroxide catalysts, which typically vary between 25 and 60 mV/dec[7], which could point towards a similar OER reaction mechanism[19].

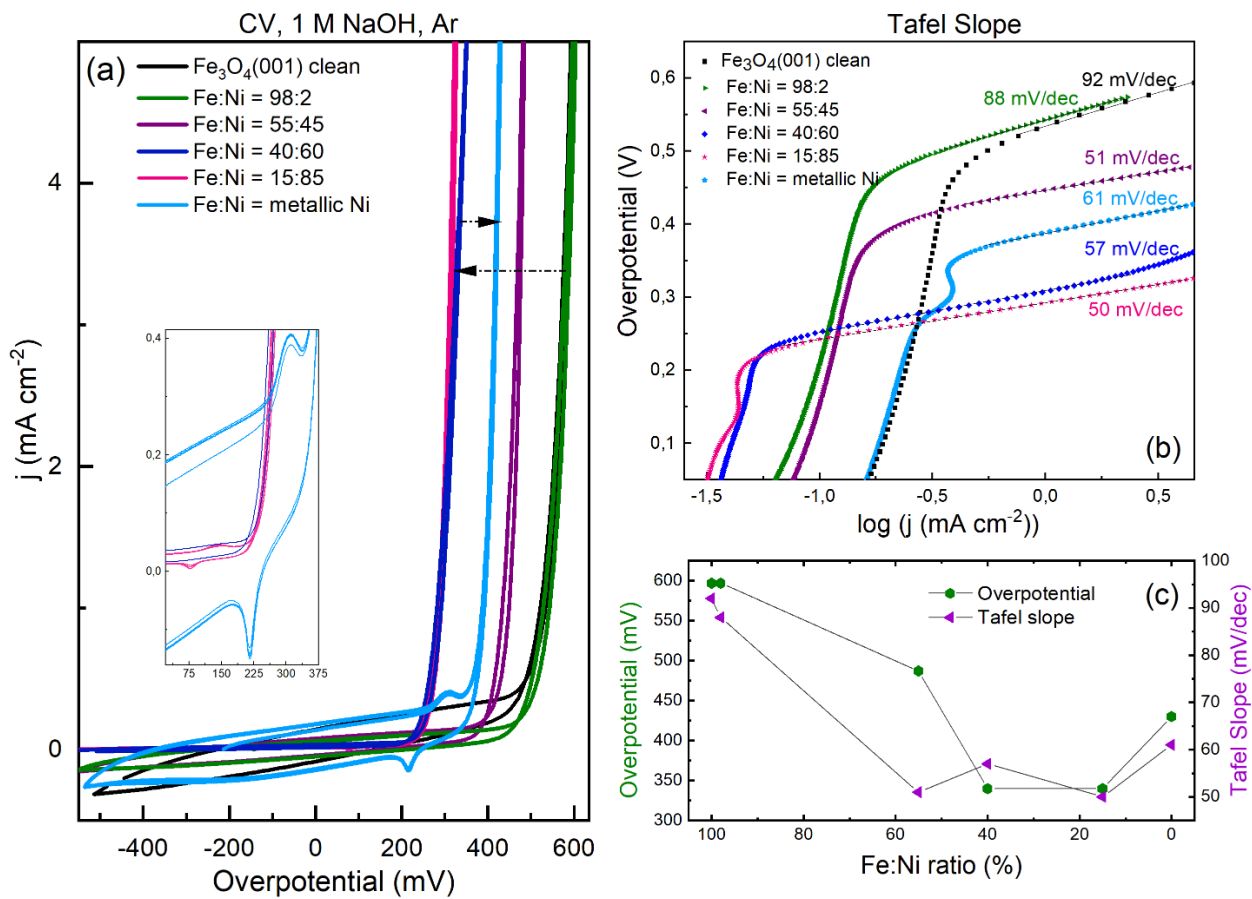

*Figure 4. Oxygen evolution reaction (OER) on clean and Ni-doped $Fe_3O_4$(001) surfaces. (a) Cyclic voltammetry acquired in 1M NaOH, Ar atmosphere, and with a 10 mV/s scan speed. (b) Tafel slopes obtained from data in (a). In these panels the clean $Fe_3O_4$(001) is plotted in black, and the Ni-doped surfaces with an Fe:Ni LEIS (surface composition) ratio of 98:2, 55:45, 40:60, 15:85, and the one covered in metallic Ni are reported in green, purple, blue, pink, and light blue, respectively. (c) Overpotential for j = 5 mA·$cm^{-2}$ (left axis) and Tafel slopes (right axis) values from (a) and (b) as a function of the surface Fe:Ni ratio (expressed in percent). This diagram shows that the lowest overpotential is obtained for an ideal surface Fe:Ni ratio between 15-40% range.*

To check whether catalyst aging in electrolyte affects the activity, we performed cyclic voltammetry on the same surfaces after leaving the Ni-doped electrodes for three days in electrolyte. Figure 5a-b show CVs of the surfaces with an Fe:Ni ratio of 40:60 (blue), 15:85 (pink), and a sample with metallic Ni clusters (light blue); the dashed curves show the performance after aging. The aged samples show a decrease of the OER overpotential by ~20-100 mV, in good agreement with what has been observed for powder catalysts prepared by wet chemistry[7,23]. Interestingly, the surface with an Fe:Ni ratio of 15:85 is similarly active to the one with Fe:Ni ratio of 40:60 when freshly prepared, but shows a much lower onset of the overpotential after aging. This observation indicates a profound structural difference in the two catalysts, despite the similar performance at first. Figure 5b shows a magnification of the capacitive regions of the CVs. The Ni-doped magnetite with metallic Ni at the surface (light blue) shows an anodic oxidation peak before OER onset and subsequent cathodic reduction in the backward scan direction. On the surface with an Fe:Ni ratio of 15:85 (pink), these peaks evolve upon cycling and aging, both in terms of charge as well as shift in overpotential (pink). However, this effect is not so marked in the case where the whole surface is covered with metallic Ni clusters, where only a (slight) shift in potential is observed (blue). The interpretation of the redox behavior is in general very difficult due to possible formation of electrically disconnected domains upon cycling because of the different conductivity of the oxidized and reduced phase[24].

Figure 5c shows the comparison of the Tafel plots for the surfaces in (a). The aged surfaces show Tafel slopes values in the range 43-62 mV/dec range, similarly to the freshly prepared catalysts (Figure 4b).

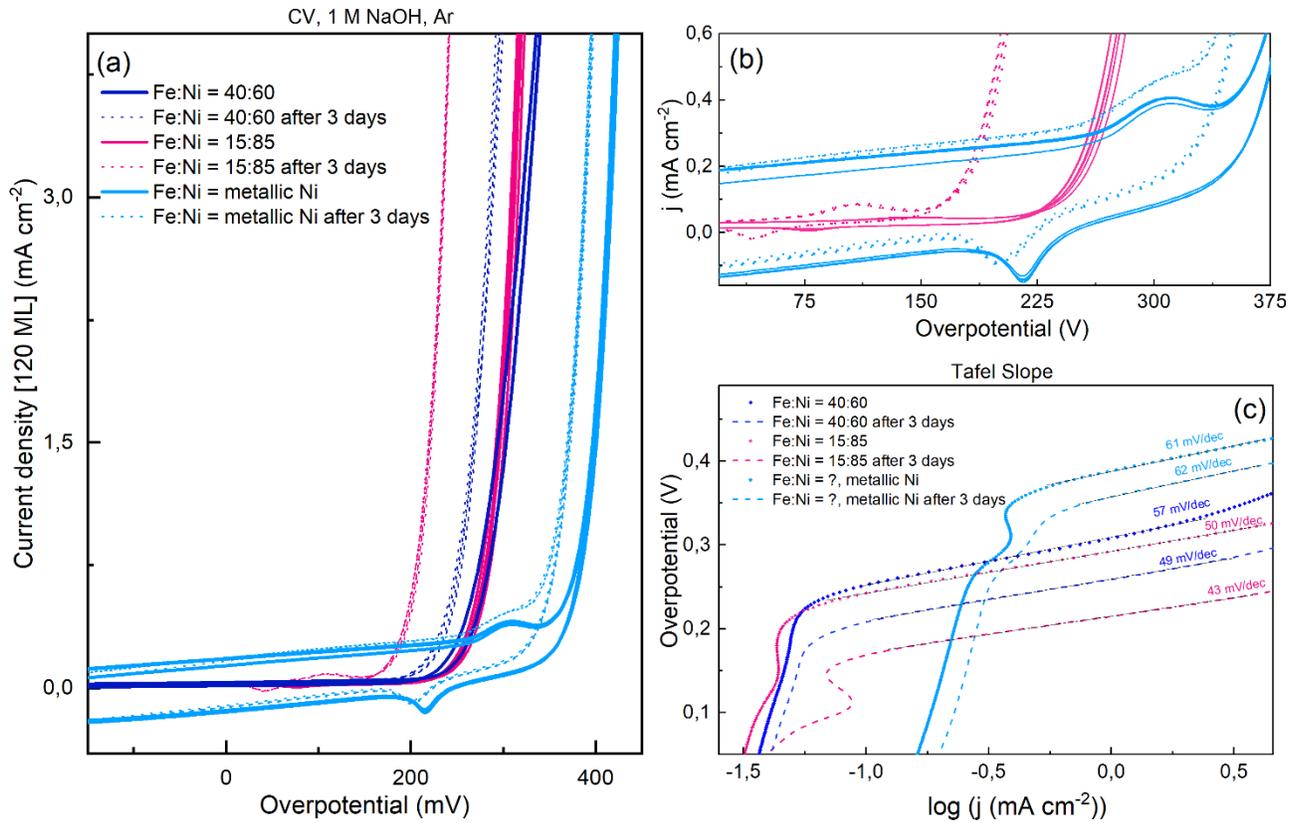

*Figure 5. Oxygen evolution reaction (OER) on clean and Ni-doped Fe$_3$O$_4$(001) surfaces before and after three days aging. (a) Cyclic voltammetry acquired in 1M NaOH, under Ar atmosphere, and with a 10 mV·s$^{-1}$ scan rate. (b) Zoom into the lower overpotential region shows changes in the anodic and cathodic waves for the surfaces with higher Ni load, before the OER onset. (c) Tafel plots obtained from data in (a) with Tafel slopes indicated. Each panel shows the activity of the Ni-doped surfaces with an Fe:Ni ratio of 40:60, 15:85, and the one with metallic Ni clusters, right after preparation (solid blue, pink and light blue), and after aging the sample in electrolyte for three days (dashed blue, pink and light blue).*

### 2.3 Catalyst Surface Characterization After Reaction

Figure 3a'-d' shows the AFM characterization of the surfaces imaged in Figure 3a-d after OER and three days aging in electrolyte. Before imaging, each surface was rinsed in milli-Q water several times, for several minutes and blow-dried using a gentle Ar flow to minimize the presence of salt residue from the electrolyte.

The morphology of the clean $Fe_3O_4$(001) remains unchanged after OER (a´), and shows a smooth appearance with the wide terraces and step bunches still visible, in agreement with earlier stability tests[18]. The presence of small particles (white) is associated with residue from the electrolyte.

Figure 3b´ shows the AFM image of the surface imaged with initially 40:60 Fe:Ni (Figure 3b) after resting in electrolyte for three days, followed by cycling the electrode until a stable current was observed (Figure 5). The terraces and step bunches remain visible underneath, but white features of irregular shape and height between 1-2 nm are now common on the surface (blue line profile in figure 3b´´).

Figure 3c´-d´ show AFM images of the 15:85 and metallic Ni surfaces, respectively, after the electrode has been exposed to the electrolyte for three days and cycled until a stable OER current was observed (Figure 5). Their morphologies appear similar in AFM. Due to the appearance of protrusions with 3-7 nm (line profile in Figure c´´, blue) and 4-8 nm high (line profile in Figure d´´, blue), it is almost impossible to discern remainders of the original surface morphology consisting of flat and wide terraces. Since the density and height of the protrusions increases with Ni content, they likely consist of a Ni-(oxy)-hydroxide phase, grown from pre-existent metallic Ni upon electrochemical cycling[23], in agreement with equilibrium potential–pH diagrams (i.e. Pourbaix diagrams) that show NiOOH as the predominant species in neutral-to-basic aqueous solutions at OER potentials[23].

### 3.4 Electrochemical Impedance Spectroscopy

Electrochemical impedance spectroscopy (EIS) measurements were performed on the Ni-doped model catalysts electrochemically investigated in Figure 4. In the OER region the EIS Nyquist plots (Figure S1a) exhibit two relaxation processes characterized by two semi-circles that can be assigned to two capacitances while the phase in Bode plots (Figure S1b) exhibits two maxima eventually merging into a broad peak. This impedance behavior is consistent with previous measurements on metal transition oxides and perovskites during the OER.[25,26,27] The EIS response can be modelled by the equivalent circuit (EC) shown as an inset in Figure 6a with a double-layer capacitance ($C_{dl}$) in parallel with the combination of a polarization resistance ($R_p$) and an adsorption pseudo-capacitance ($C_{ads}$) in parallel with a resistor $R_s$. The $C_{dl}$ element accounts for the charging of the electrified interface. $C_{ads}$ models the accumulation of an adsorbed intermediate involved in the rate-limiting step of the OER. The sum of the resistive elements $R_s$ and $R_p$ bear a physical meaning as the zero-frequency electron transfer resistance defined as $R_f = R_p + R_s$, i.e., the slope of the steady-state polarization curve after Ohmic-drop compensation. $R_\Omega$ represents the electrolyte resistance. It has to be noted that both capacitors were modeled as constant phase elements (CPEs), defined as $Z = C_{n=1}^{-1}(j\omega)^{-n}$, where $C_{n=1}^{-1}$ is the impedance of the capacitor without frequency dispersion, i.e., if the coefficient *n* = 1 which is the case for a perfect capacitor. The interpretation of the CPEs dispersion coefficient *n* is varied and complicated; its origin has been attributed to surface roughness, inhomogeneities, or inhomogeneous adsorption of ions[28]. In the double-layer region, prior to the onset of the OER, we will show in a separate work that the impedance response of the single-crystal magnetite electrode has to be modified by adding a Warburg element in series with $C_{ads}$ corresponding to a diffusion impedance that we attribute to electrolyte cations intercalating into the oxide (Figure S2a). Of interest in this work is the impedance behavior in the OER region.

All the surfaces investigated in this work, with the exception of the one fully covered by metallic Ni clusters (light blue), show constant double layer capacitance values in the 10-25 $\mu F \cdot cm^{-2}$ range prior to the OER onset (Figure 6a). The exponent of the CPE element used for the fitting was equal to 1 in the double-layer region (Figure S2d) and diverged from 1 at high current densities or when Ni is exposed such that $Ni(OH)_2$ is oxidized to NiOOH. These values are comparable to a $C_{dl}$ observed on metallic single crystals, suggesting that our catalysts have a perfect capacitor-like behavior. Figure 6c shows the value of this capacitance as a function of the Ni content: $C_{dl}$ slightly increases from 10 to 15 $\mu F\ cm^{-2}$ as the Ni loading increases, but a higher value is observed in the case of the surface fully covered with Ni metallic clusters (180 ML). The higher $C_{dl}$ values observed for this surface may be explained by the formation of an irregular $Ni(OH)_2$ layer upon oxidation of the metallic Ni by contact with the electrolyte. In this way, more active surface area is exposed to the electrolyte and polarized, leading to a higher $C_{dl}$.

The adsorption capacitance plot in Figure 6c shows that the Ni-doped $Fe_3O_4$(001) surfaces display a peak with similar $C_{ads}$ values independent of the Ni doping level, which however shifts to lower overpotential as the Ni load increases (Figure 6d). The surface fully covered with metallic Ni clusters appears to develop two additional capacitance peaks (Figure 6c). The overlay of the corresponding CV and $C_{ads}$ in Figure S3e, shows that the additional initial (pre-)peak is observed at the same potential as the $Ni(OH)_2$ oxidation peak.

The group of Bandarenka[29,30] associated the observation of peaks or increase in $C_{ads}$ to the adsorption of OER reaction intermediates and reported them for various transition metal oxides. These observations suggest that the formation of the intermediate species before the onset of the OER involves similar mechanisms for pure and Ni-modified magnetite. This is also supported by the fact that value of $C_{ads}$ retains similar values at the maximum of the peak. From the capacitance data in Fig.6a and c we can draw the following conclusions: (i) given that the initial $C_{dl}$ values hardly varies with Ni loading, there is no significant increase in electrochemically active surface area, and the catalytic effect of Ni shown in Fig.4 cannot be ascribed to an effective enhancement of the surface area; (ii) the fact that a similar peak in $C_{ads}$ is observed for all surfaces, also the one where Fe is expected to be the only active site (98:2), would be in agreement with the commonly held view that Fe is the active site in NiFe catalysts, but that it becomes more active in an Ni environment. The presence of two peaks in the EIS of the metallic Ni-decorated surface if not a noise effect can be interpreted as two types of adsorbates on Ni (and perhaps Fe) sites that are accessible due to the porosity and layered structure of Ni films that allow access to active sites down to 5 nm in depth[31].

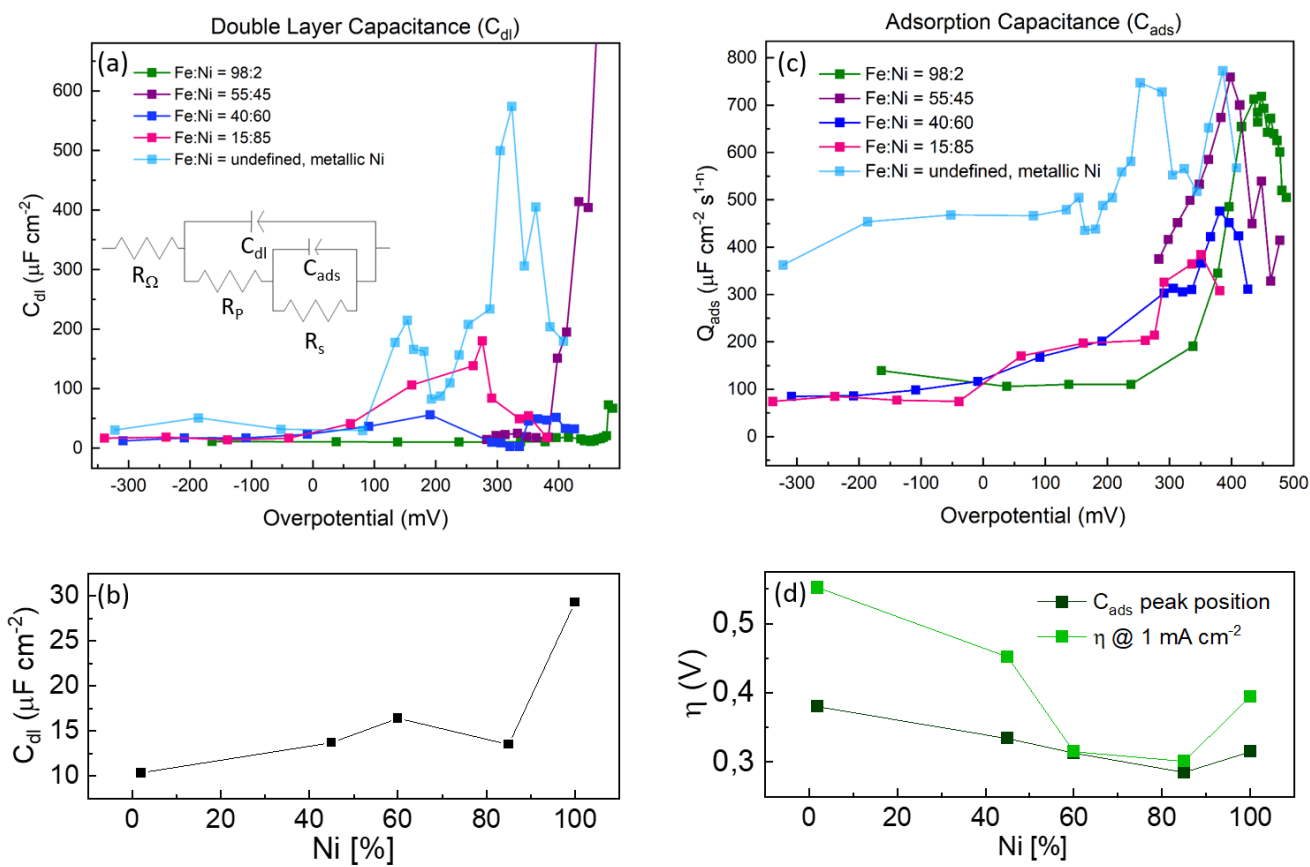

*Figure 6. Electrochemical impedance spectroscopy on Ni-doped $Fe_3O_4$(001) surfaces. (a) Double layer capacitance and (b) minimum values of $C_{dl}$ in the double-layer region;(c)) adsorption capacitance ($C_{ads}$) and potential of the maximum of $C_{ads}$ for the Ni-doped magnetite surfaces. In each panel of a) and c) the Ni-doped $Fe_3O_4$(001) surfaces with an Fe:Ni ratio of 98:2, 55:45, 40:60, 15:85, and the one fully covered in metallic Ni clusters are reported in green, purple, blue, pink, and light blue, respectively. The equivalent circuit used to fit the electrochemical impedance spectroscopy data is shown as an inset in (a) and it*

*consists of the following elements: $R_\Omega$ represents the uncompensated solution resistance; $C_{dl}$ models the double layer capacitance; the polarization resistances $R_P$ in combination with $R_S$ are commonly interpreted as the charge transfer resistances of the electrosorption and desorption processes. Finally, $C_{ads}$, the adsorption pseudo-capacitance in parallel with $R_S$ models the relaxation of the charge associated with the adsorbed intermediate*[25].

## 4   Discussion

The experimental data acquired on clean and Ni-doped $Fe_3O_4$(001) surfaces show that Ni doping enhances the OER activity of magnetite. Electrochemical voltammetric responses, in combination with surface sensitive techniques, suggest a strong dependence of the OER activity from the atomic structure of the surface exposed to electrolyte. In particular, LEIS measurements indicate that the catalyst with the best OER performances, with an overpotential of 340 mV vs RHE at 5 mA·cm$^{-2}$, exhibits a surface Fe:Ni ratio of 40:60.

In order to shed light on how the presence of Ni affects the magnetite atomic surface structure-activity relationship, the following observations have to be considered:

We have previously shown that following 1 ML Ni-doping and subsequent mild annealing at 200 °C, the Ni atoms fill all the vacant sites in the $Fe_3O_4$(001) subsurface, resulting in neighboring Fe and Ni in the second surface layer[8]. The voltammetric response of this surface (green, Fig. 4a) shows no improvement in the OER activity compared to the clean magnetite. Corresponding LEIS measurements (see supplemental material, Figure S4) suggest that this surface exhibits a surface Fe:Ni ratio of 98:2, confirming that almost no Ni is present in the outermost surface layer. These results suggest that the presence of subsurface Ni is insufficient to improve the OER activity. Based on our XPS and LEED data, we propose that modification of the $Fe_3O_4$(001) surface with a Ni load > 1 ML leads to the formation of a multilayer mixed ferrite spinel oxide with a structure similar to $Ni_xFe_{3-x}O_4$-like systems. Now the model catalyst exposes both, Fe and Ni atoms in the outermost surface layer as seen from LEIS. In these conditions, the OER activity increases, reaching a maximum when the surface exposes an optimum Ni content of 60-85%. XPS and LEED suggest that the structure of this surface stays characteristic of mixed spinel up to a Ni doping corresponding to a surface Fe:Ni ratio of 40:60. Higher Ni doping results in the formation of metallic Ni clusters, which compromise the spinel long-range order, leading to a loss of atomic control without further enhancement of the activity and, eventually, a decrease in the OER activity.

Our AFM results suggest that the surface prepared with an Fe:Ni ratio of 40:60 appears stable after OER, albeit with some new features, 1-2 nm high, scattered all over the surface. In contrast, the surfaces with higher Ni loads show the growth of a new phase, which increases in volume and roughness (effective surface area) as the metallic Ni concentration increases. This suggests the growth of a new phase on top of the doped magnetite surface. On the basis of our XPS results, as well as earlier studies[23,32,33], we interpret this phase as the growth of Ni-(oxy)-hydroxide. Similar phases have been also observed on powder Fe-Ni based catalysts, and have reported in literature to affect the catalytic activity towards OER[7]. In particular, Burke et al.[23] observed that electrochemical cycling leads to a transformation from nano-crystalline $NiO_x$ films to a layered (oxy)-hydroxide that correlates with an increase in OER activity. Similarly, Deng et al.[33] monitored the dynamic changes of single layered $Ni(OH)_2$ using in situ electrochemical-AFM, and observed dramatic morphology changes already after one linear voltammetry sweep, as well as a direct relation between increase in OER activity and increase in volume and redox capacity of the layered oxy-hydroxide phase. Our results are, however, are not entirely in agreement with these observations. The increase in volume and surface area of the hydroxide phase does not correlate directly with our catalysts' activity: the surface with the highest amount of (oxy)-hydroxide phase and redox capacity is ≈ 200 mV less active than the (almost) flat surface with Fe:Ni ratio 40:60. Based on these observations, we can rule out the possibility of the layered Ni-(oxy)-hydroxide as the active phase in our catalysts.

It is also important to mention that the activity exhibited by the surface prepared with a Fe:Ni ratio of 40:60, with an overpotential of 340 mV vs RHE is comparable to values reported for OER on (Fe)Ni based catalysts[34-40]. For comparison, the overview in Table 1 shows a selection of some of the best OER catalysts based on Ni-Fe oxides reported in literature. The lowest overpotential values measured on these catalysts at 5 mA·cm$^{-2}$ vary typically in the 210 - 347 mV vs RHE range (in 1 M KOH or NaOH electrolyte). Similar overpotential values were also obtained from our surface prepared with a higher Ni load (Fe:Ni = 15:85). When comparing the latter surface with the one having an Fe:Ni ratio of 40:60 after electrochemical cycling and subsequent aging for three days in electrolyte, a different activity trend is observed. On the one hand, both surfaces show a significant increase in activity following voltammetric cycling and aging, in agreement with previous studies[3,33]. On the other hand, their activity does not increase in the same way. Surprisingly, the surface with metallic Ni shows a much lower onset of the OER overpotential than the one with an Fe:Ni of 40:60 , despite the similar performances when freshly prepared. This surface is by far the most active with an overpotential of 247 mV vs RHE. However, it has to be taken into account that this surface, being characterized by the presence of metallic Ni, does not show a well-defined spinel structure and, therefore, cannot serve as a model system. Since one of the scopes of this work is to propose a working model system for the understanding of the OER mechanism, a compromise between activity and the ability to preserve atomic control has to be made. In this regard, the surface with a Fe:Ni ratio of 40:60, very well defined and highly active, fits the criteria to be used as model catalyst.

| Catalyst | Ni:Fe ratio | Electrolyte | Overpotential vs RHE (mV) at $j = 5$ mA·cm$^{-2}$ | Tafel Slope (mV·dec$^{-1}$) | Ref. (year) |
|---|---|---|---|---|---|
| Ni-Fe$_3$O$_4$(001) | 60:40 | 1 M NaOH | 340 | 57 | This work |
| NiFeO$_x$ film | 9:1 | 1 M KOH | 336 | 30 | Ref.[34] (2013) |
| NiFe LDH | 3:1 | 1 M KOH | 347 | 67 | Ref.[35] (2014) |
| NiFe LDH | 80:20 | 1 M NaOH | 260 | 21 | Ref.[36] (2016) |
| Ni-Fe | 4:1 | 1 M KOH | 331 | 58 | Ref.[37] (2015) |
| NiFe LDH | 78:22 | 1 M KOH | 280 | 47.6 | Ref.[38] (2014) |
| NiFe LDH/CNT | 5:1 | 1 M KOH | 247 | 31 | Ref.[39] (2013) |
| NiFe/NF | 3:1 | 1 M KOH | 215 | 28 | Ref.[40] (2015) |

*Table 1. Comparison of our model system activity with real catalysts from recent works.* Abbreviations: LDH = layered double hydroxide, NF= Ni foam, CNT= carbon nanotubes.

Finally, the analysis of the Tafel plots and adsorption capacitance measurements can help extracting information to identify the OER active sites. Our Ni-modified magnetite surfaces show similar absolute Tafel slopes values (Figures 5b and 6c) in the 43-62 mV·dec$^{-1}$ range, independently of the degree of Ni doping for Fe:Ni ratios down to 15:85. Furthermore, the clean and the Ni-modified surfaces shows similar maximum values of the adsorption capacitance before the OER onset. These values are associated to the appearance of the OER precursors[29,30] and the shift to lower overpotentials as the Ni doping increases and finally reaches a steady value whit the optimal Ni content (Ni content ≈ 60-80%).

To explain these observations, we propose the following scenario: the intermediate species that forms on the surface before the onset of the OER might be the same on the clean surface as well as on the Ni-modified one, indicating Fe as the active sites. Accordingly, the right amount of Ni in the spinel surface does not cause the intermediates formation but facilitates it. Similar conclusions have been proposed by Bell and co-workers who used DFT to compared the OER activity of pure and Fe-doped γ-NiOOH and of pure and Ni-doped γ-FeOOH catalysts[41]. They showed that pure γ-NiOOH adsorbs the OER intermediates too weakly and pure γ-FeOOH too strongly. They found a considerable increase in activity for Fe sites that are surrounded by Ni next-nearest neighbours in both γ-NiOOH and γ-FeOOH. Similar results have also been obtained by Klaus et al. who, on the basis of turnover frequencies (TOF) calculations, proposed Fe atoms as the OER active sites in Fe-doped NiOOH catalysts[22].

Nevertheless, despite the OER mechanism on NiFe based catalysts is still unclear together with several fundamental open questions such as the clear identification of the rate limiting step intermediate, our results tend to confirm that the OER intermediates are located on Fe sites, the surrounding Ni having a promoting effect on the latter. Furthermore, a deep understanding of the observed electrode aging effect on the OER activity, following electrochemical cycling and long exposure of the material to the electrolyte, remains open, but reveals the importance of the nature of the electrolyte and its interaction with the material.

Moreover, it should be pointed out that the use of a single crystal enables an accurate determination of the electrochemically active surface area (ECSA) of these materials and provides reference values for the double-layer capacitance and adsorption capacitance on Fe-Ni based catalysts. The double-layer capacitance values are slightly affected by the Fe:Ni ratio and this should be taken into consideration for further determination of the ECSA of such electrodes[29,30]. Additionally, our results point out that, beyond the Fe:Ni ratio, the nature of the interface (spinel or separated NiOOH/Fe-Ni spinel) significantly affects the capacitance of the interface and its use as a reference for ECSA determination could be compromised.

**Conclusions**

The high intrinsic OER activity of mixed Fe-Ni oxides motivated our efforts to make further steps in the understanding of the fundamental roles of Fe and Ni in OER catalysis.

In this work, we show a combined surface science/electrochemistry approach for the preparation of well-defined Ni-modified $Fe_3O_4$(001) surfaces and the investigation of their electrochemical performances with respect to OER. We have found that the surface prepared with an Fe:Ni ratio of 40:60 shows performances comparable to those of the best powder catalysts reported in literature, and still maintains a well-defined structure. Being at the same time simple, highly active, and stable under operation conditions, this surface is an ideal candidate to serve as a working model system to gain atomic-scale insights into the complicated OER mechanism. Electrochemical impedance spectroscopy enables us to confirm that on our Fe-Ni catalyst, the active site for the OER is located on Fe atoms at the surface regardless of the Ni:Fe ratio in the structure.

Putting our results in the context of future perspective, a well-defined model system such as the Ni-modified $Fe_3O_4$(001) presented in this work is desirable to address the fundamental aspects that are still controversial. With a limited variety of possible adsorption sites and being accessible to methods benefitting from on single-crystal surfaces, this model surface could thus be used for further investigations on the exact nature of the adsorbates involved in the rate limiting step, using *in-situ* surface science techniques, to shed more light on key parameters to improve the stability and activity of amorphous catalysts used in water splitting devices. We also believe that the good agreement of our results with what reported in the literature for powder or amorphous catalyst makes our model surface worthwhile to be used as a model to guide future computational studies.

**Acknowledgements**

This work was supported by the European Union under the A-LEAF project (732840-A-LEAF), by the Austrian Science Fund FWF (Project 'Wittgenstein Prize, Z250-N27), and by the European Research Council (ERC) under the European Union's HORIZON2020 Research and Innovation program (ERC Grant Agreement No. [864628]).